# The oxDNA coarse-grained model as a tool to simulate DNA origami

Jonathan P. K. Doye, Hannah Fowler, Domen Prešern, Joakim Bohlin, Lorenzo Rovigatti, Flavio Romano, Petr Šulc, Chak Kui Wong, Ard A. Louis, John S. Schreck, Megan C. Engel, Michael Matthies, Erik Benson, Erik Poppleton and Benedict E. K. Snodin

**Abstract**
This chapter introduces how to run molecular dynamics simulations for DNA origami using the oxDNA coarse-grained model.

**Key words** DNA origami, molecular simulation, coarse-grained models

**1 Introduction**
DNA origami provides an attractive approach for designing structures and devices on the nanoscale. Particular benefits include the relative ease of the design and assembly processes, the addressability of the resulting structures, and the fine structural control that is achievable. These features \are some of the reasons that the field of DNA origami has seen spectacular growth since its inception in 2006 [1], and ever more complex (for example, in terms of size [2], function [3] and programmable motions [4]) origami designs are being realized.

Being able to model the properties of DNA origami has the potential to contribute significantly to the future of this field. Potential benefits include (i) a more detailed view of origami structure than is typically available from experiment, (ii) a realistic picture of the effect of thermal fluctuations on origami structure and behaviour (as opposed to the more static viewpoint inherent to design programmes), (iii) the ability to pre-screen the properties of putative origamis prior to experimental realization, and (iv) the ability to identify the physical causes of observed behaviours and thus to contribute to a rational design process. Furthermore, access to these types of insight is becoming more important as the desired functional complexity of the origami increases and the design process becomes more challenging.

Modelling can be performed at a variety of levels of resolution with inevitable trade-offs between the detail available and the time scales required for computation. At the coarser end are models like CanDo [5,6] and mrDNA [7] in which origamis are represented as a series of mechanical elements (e.g. duplexes, single strands, junctions) with known properties, where the ease of use, robustness and short computation times have led to widespread usage. At the other end are atomistic simulations where all atoms of the origami and local solution environment are represented [8]. In the middle are models like oxDNA [9,10,11], the focus of the current chapter. OxDNA is a nucleotide level model of DNA where the interactions illustrated in Fig. 1 have been fitted to capture well the structure and mechanics of duplex DNA (e.g. bend and twist persistence lengths) and single-stranded DNA (e.g. the force-extension curve) and the thermodynamics of hybridization. These features make oxDNA very generally applicable, and allow it to realistically describe the many origami properties that are outworkings of these basic biophysical properties of DNA. In terms of structure, for example, it is able to describe the local splaying out of helices at four-way junctions in origami [12], the overall global structure of the most accurately determined origami structure to within the experimental resolution [12], the response of origamis to internal stresses that lead to an elastic response (e.g. twisting [11,13] and bending [14]) or to coupling to internal degrees of freedom (e.g. breaking of base pairs, unstacking at nicks and junctions), the properties of single strands as linkers in flexible origamis [4,15,16] or as bearers of tension [14,17]. In terms of mechanics, it can describe the elastic properties of origamis, the yielding of origamis under tension [18], and even the chiral helicoidal fluctuations of twisted DNA nanotubes [13]. It can also be used to describe the thermodynamics and dynamics of hybridization processes associated with DNA origami, e.g. self-assembly [19] and the actuation of origami devices [20].

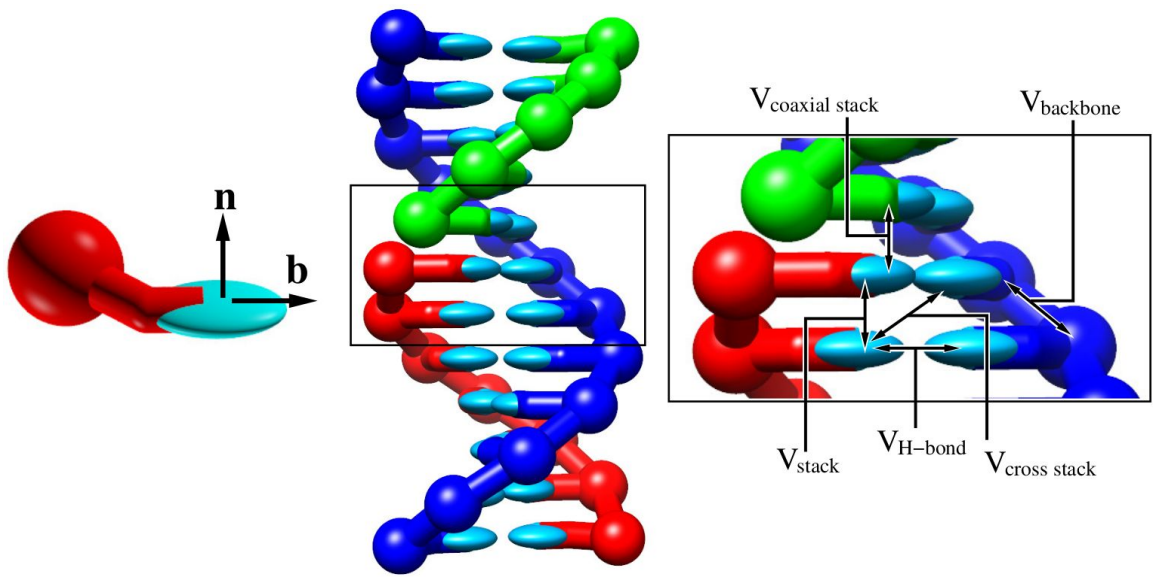

**Fig. 1** The oxDNA model. Each nucleotide is a rigid body with sites corresponding to the centres of the different interactions. The position and orientation of a nucleotide are defined by **r** the position of the notional centre of mass, **b** a "base" vector collinear with the stacking and hydrogen-bonding sites and **n** a vector normal to the notional plane of the base. The basic interactions in the model are the (FENE) backbone potential connecting backbone sites, a hydrogen-bonding potential between complementary nucleotides, (coaxial) stacking interactions between bases that are (non-) adjacent along the chain, electrostatic repulsion between backbone sites, and cross-stacking interactions between bases that are diagonally opposite in the duplex.

However, as with any coarse-grained models, there are features that cannot be fully described. Here, we wish to provide some caveats about the model, along with the possible implications for origami modelling. Firstly, isolated four-way junctions in their stacked form in oxDNA have a left-handed twist angle between the two helices [12], whereas experiments indicate that the preferred form is right-handed.

For origami modelling this is unlikely to be an issue as most such junctions are constrained to adopt an anti-parallel configuration. Secondly, the extensional modulus in oxDNA is significantly higher than for actual DNA [9]. In the parameterization of the model it did not prove possible to generate a model that could simultaneously be a good fit of the bend, twist and extensional moduli, and a choice was made to prioritise accurate modelling of bend and twist [9]. This shortcoming may affect mechanical responses of origami that couple to the stretching of individual helices.

Thirdly, the electrostatic interactions included in the model are of a relatively simple Debye-Hückel form that has an explicit dependence on the ionic strength of the solution. This has been fitted to reproduce the [$Na^+$] dependence of the hybridization thermodynamics for [$Na^+$]>0.1M. Such a simple description clearly cannot capture ion-specific effects. For example, $MgCl_2$ has an ionic strength that is just 3 times that of the ionic strength of NaCl, but has a much greater effect than this on both duplex and origami stability. Following Rothemund's original protocol [1], DNA origami are often assembled at [$MgCl_2$]=12.5 mM, whereas very high NaCl concentrations are required for origami assembly [21]. This difference is partly due to the particular stabilization of the stacked form of four-way junctions by $Mg^{2+}$. Although in oxDNA the transition from the open to the stacked form of four-way junctions occurs at too low a $Na^+$ concentration, this is probably helpful for modelling of origamis, as the junctions consequently behave more like in the typical $Mg^{2+}$ conditions used in experiments. We recommend using [$Na^+$]=1M in oxDNA as representative of those conditions.

Fourthly, due to the relative lack of high-quality physical chemistry data, the parameterization of the coaxial stacking interaction in oxDNA should be viewed with some caution. Currently, the interaction has no sequence dependence, and its magnitude represents a compromise between experimental data that could not be simultaneously reproduced [11]; further, little is known about the orientation dependence of this interaction. This shortcoming may affect the modelling of blunt-ended stacking of helices that is being increasingly used to mediate multi-origami assembly [2,22]).

Currently, the oxDNA model is available as its original stand-alone simulation programme, which can be used for both Monte Carlo and molecular dynamics, and also through LAMMPS [23], a widely-used molecular dynamics package. Parallelization is much more straightforward for molecular dynamics than for Monte Carlo, and for this reason molecular dynamics is the favoured approach for simulating large structures such as DNA origami.

Currently, there are two main versions of oxDNA, the original [9] and a second version [11] (sometimes called "oxDNA2" and specified by `interaction_type = DNA2` in the input file). For origami simulations one should always use oxDNA2, as its properties (e.g. DNA pitch, twist at junctions and nicks) have been fine-tuned to match experimental data on DNA origami [11]. Other additional features of oxDNA2 include major-minor grooving and electrostatics. The parameter sets for the models also come in sequence-averaged and sequence-dependent [10] varieties. In the sequence-averaged model, the interaction strengths are independent of the identity of the base (although of course base pairing can still only occur between Watson-Crick pairs). In the sequence-dependent model the interaction strengths have been tuned to reproduce the sequence thermodynamics of hybridization, but sequence-dependent structure and mechanics have not been explicitly incorporated. Generally, we use the sequence-averaged model to study the general behaviour of the DNA, and the sequence-dependent model when comparing to a specific experimental system or when we are specifically interested in the sequence dependence of the behaviour. Note, there is also an RNA-equivalent of the oxDNA model, which is called oxRNA [24] and has been parameterized in a similar manner, i.e. with a focus on reproducing the thermodynamics of RNA, and which is particularly useful for RNA nanotechnology.

## 2 Materials

### 2.1 Software

1. **oxDNA.** oxDNA is used to refer both to the coarse-grained model of that name and its dedicated simulation code. The oxDNA simulation code can be downloaded from https://sourceforge.net/p/oxdna/. Documentation (somewhat in need of updating) is currently at https://dna.physics.ox.ac.uk. Installation is known to work on Linux and Mac OS X.
2. **tacoxDNA** is a webserver (http://tacoxdna.sissa.it/) that provides a user-friendly interface to interconvert different DNA file formats (including those used by the most popular DNA design tools) in order to facilitate simulations with oxDNA. The standalone Python scripts are also available. A full description of tacoxDNA's capabilities is available in Ref. [25].
3. **oxView** is a browser-based viewer that allows oxDNA configurations and trajectories to be visualized and manipulated. It is also integrated with a package of Python analysis tools that provides for many common simulation analysis needs. See https://github.com/sulcgroup/oxdna-viewer and https://github.com/sulcgroup/oxdna_analysis_tools. A full description of the capabilities of oxView and its associated analysis tools is available in Ref. [26].
4. **cogli1** is a convenient lightweight viewer that can directly read in oxDNA configurations. It can be downloaded from https://sourceforge.net/projects/cogli1/. Calling cogli1 without any options lists the options available and the keyboard and mouse bindings. Note that, although it has been made publicly available, it is primarily a research tool for the developers and their collaborators.

### 2.2 Files

Files for the examples considered in this chapter can be obtained as a zip file both from the publisher's website and from the Oxford University Research Archive (https://deposit.ora.ox.ac.uk/datasets/uuid:7a111527-3c1a-4c0f-af89-774b01f43abd). The input files for oxDNA, named `input_min`, `input_relax` and `input_sim`, contain the simulation parameters (number of time steps, salt concentration, temperature, etc.) as well as the paths to the input and output for the simulation. They can be read and edited in any text editor.

## 3 Methods

Here we will illustrate how to simulate DNA origami using oxDNA for three examples. The first example is an asymmetric "pointer" origami block whose structure has been determined to high accuracy by cryoEM [27]. The second is a six-helix bundle that is designed to have two turns of left-handed twist [28].

The third is a "switch" that has two arms and can adopt open and closed states [22]. In the open state, which we study here, the two arms, which are connected by short single-stranded linkages, can rotate relatively freely with respect to each other. In the closed state, the two arms interlock and are held together by blunt-ended stacking between helices. The files for these examples are in the directories/folders "pointer", "2xLH", "switch". In these case studies we will perform the simulations using the native oxDNA simulation code rather than the implementation in LAMMPS, because only the former can be currently run on a GPU. Once GPU support is available for oxDNA in LAMMPS, this code could provide an appropriate alternative (note, tacoxDNA provides the means to interconvert between the formats required for the oxDNA simulation code and for LAMMPS).

For convenience, the oxDNA simulation code uses its own internal unit system (often referred to as simulation units). The values for parameters in the input files typically have to be given in these units. Similarly, input and output configurations are also in simulation units. The conversion factors are given in **Note 1**.

Note that, although the following is written as commands entered via the command line, in practice, one will generally want to run the relaxation and simulation stages on a cluster and to submit the jobs via a queuing system rather than interactively from the command line.

### 3.1 Conversion to oxDNA format

This chapter will not cover how to design an origami, but rather we assume that there is a design file available for the origami of interest that has been produced by one of the relevant computer-aided design programs. The first task is to take the design file and convert it into an initial configuration in the oxDNA format. tacoxDNA can be used to achieve this conversion, either using the interface on the web-site or the accompanying suite of Python scripts. Currently, it can convert from cadnano [29], Tiamat [30], CanDo [6] and vHelix [31] formats. Also, some of the more recent design tools also allow one to directly output into oxDNA format. E.g. vHelix, Adenita [32] and magicDNA. TacoxDNA can also convert an all-atom `.pdb` structure to oxDNA format; this is particularly useful for conversions from the suite of design programmes for wireframe structures developed in the Bathe group [33,34,35] (*see* **Note 2**). These tools allow oxDNA configurations to be generated for all the most popular DNA nanotechnology design programmes. The conversion tools generally output two files: an oxDNA configuration file (`.oxDNA, .dat` or `.conf`) specifying the positions and orientations of the nucleotides and an oxDNA topology file (`.top`) specifying the sequence and which nucleotides are covalently bonded to which along the DNA backbone.

In our three examples the origami designs have been produced by cadnano [29], and the cadnano `.json` files are in the relevant directory of the files. When using tacoxDNA, the cadnano lattice used has to be specified. It is a square lattice for the pointer, and a hexagonal lattice for the six-helix bundle and the switch. Once converted, the configuration and topology files should be added to the relevant directory (*See* **Note 3**).

The initially converted geometries are illustrated in Fig. 2. These figures have been produced using cogli1. A typical command to launch cogli1 and load the oxDNA configuration is

```
cogli1 -m -v -t xxxx.top xxxx.conf
```

To view configurations in oxView, one need only to drag and drop the configuration and topology files into the browser window running oxView. Note, the tacoxDNA webserver provides a direct link to view the converted cadnano files using oxView. In addition, oxView allows movies of trajectories to be easily created.

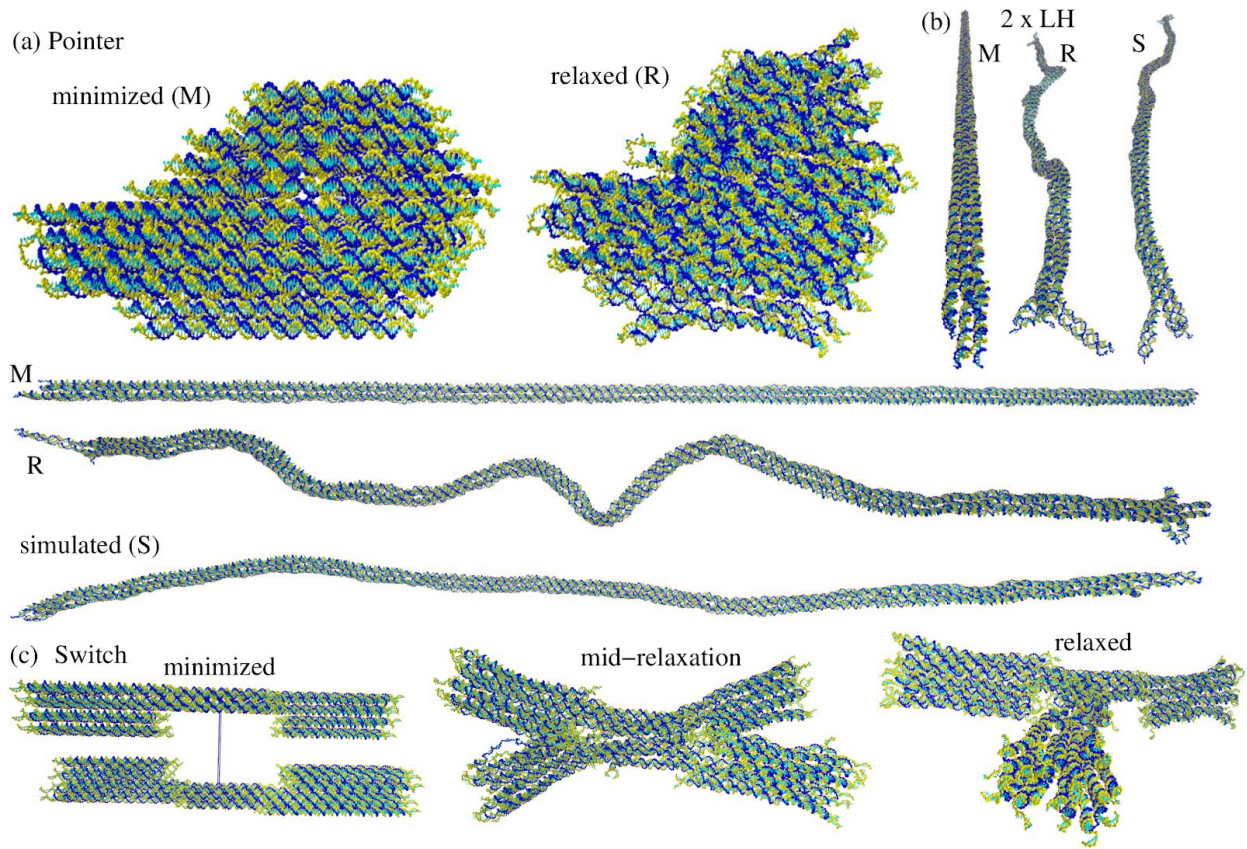

**Fig. 2** Images of the three origami systems (a) pointer, (b) 2xLH and (c) switch after minimization (M), relaxation (R) and simulation (S).

### 3.2 Relaxation of initial geometry

The configurations produced by the above conversion are typically not an appropriate starting point for a standard molecular dynamics simulation run, as there are usually some nucleotides whose excluded volumes overlap somewhat and maybe some backbone bonds that are too long. These give rise to extremely large forces that would cause the molecular dynamics simulation to fail. Therefore, it is first necessary to "relax" the structure to remove the above issues. We typically do this relaxation in two stages. (Note that for simple origamis that have a realistic starting configuration the second stage is not always necessary. It is required for the three examples.)

In both stages, we use a modified backbone potential. The standard oxDNA backbone potential is a FENE potential. This potential diverges beyond a certain distance (approximately 0.873 nm). The purpose of the modified potential is both to remove this divergence and to ensure that the forces resulting from stretched

bonds do not damage (e.g. by breaking base pairs) the origami structure. The modified potential has the form:

$$V_{mod} = V_{FENE} \qquad \text{for } r \leq r_{max}$$
$$V_{mod} = A\,r + B\,\log(r) + C \quad \text{for } r > r_{max}$$

where $V_{FENE}$ is the original backbone potential, $r_{max}$ is the distance at which the force due the FENE potential is equal to $F_{max}$ (the value of $F_{max}$ is set in the input file using `max_backbone_force`). $A$ corresponds to the limiting value of the force at large $r$ and is set using the variable `max_backbone_force_far` in the input file; $B$ and $C$ are chosen so that the modified potential is continuous and differentiable at $r_{max}$. The original and modified potentials are illustrated in Fig. 3 along with the resulting force. The modified force increases relatively gently as $r$ increases with the force tending to $A$ at sufficiently large $r$.

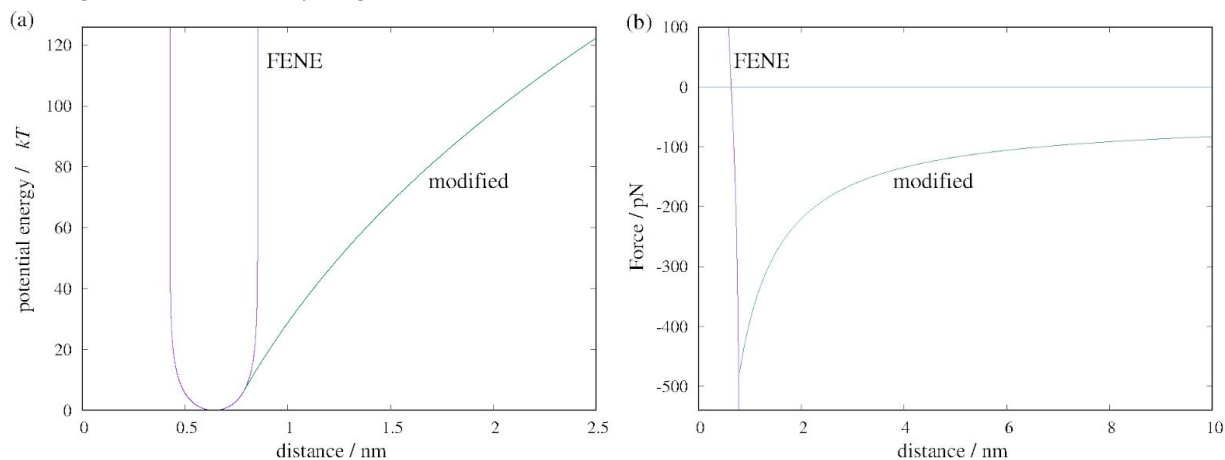

**Fig. 3** (a) The FENE backbone potential and its modified form at long range for parameter values $F_{max} = 10$; $A = 1$ (values in simulation units; equivalent to 486 pN and 48.6 pN). (b) The forces due to these potentials.

The first stage involves running a minimization algorithm for a few thousand steps (this is specified by setting `sim_type = min` in the input file) (*see* **Note 4**). This is usually sufficient to remove all particle overlaps, and those overstretched bonds that are too long by only a relatively small amount (on the order of a few nanometres). It can also help nucleotides that are designed to be base-paired to correctly orient themselves to fully base-pair. The minimization runs on a single CPU core, but even for a full-size origami it should take no more than a few minutes.

To run the minimization enter the command
        `oxDNA input_min`
from the relevant directory. The above assumes that the oxDNA executable is in one's path. If not, the full location of the executable should be given.

The standard set of files created by oxDNA provide the energy, trajectory, and final configuration. In the supplied input files, we have chosen them to be named `energy_min.dat`, `trajectory_min.dat` and

`last_conf_min.dat`. The configurations in the trajectory file can easily be viewed using cogli1 or oxView. Typically, the changes in structure will be very small and barely noticeable.

The second stage involves a fairly standard molecular dynamics simulation but with the modified backbone potential. This can be run on a GPU (*see* **Note 5**), and its aim is to allow relaxation that requires larger-scale motions. For a design file that provides a pretty accurate representation of the three-dimensional structure of the origami (e.g. the pointer), the first stage is virtually sufficient, and this stage can be relatively short. The two other case studies both require larger-scale motions. The initially converted geometry of the six-helix bundle is untwisted, but has significant internal stresses (due to the "deletions" in the design [36]) that are partially resolved by the global twisting of the structure. The switch has two sections that are able to freely rotate with respect to each other and are connected by short single-stranded sections. These are significantly extended in the starting structure and the two blocks are in an atypical parallel arrangement (Fig. 2).

To run the relaxation simulation simply enter (*see* **Note 6**):
```
oxDNA input_relax
```

In these examples, the lengths of the runs are different, reflecting the differing extents of the motions required to achieve relaxation. For the pointer we only use $10^4$ steps (where we use an integration time step of 0.005 (in simulation units)). The six-helix bundle requires on the order of $10^5$ steps. The switch requires on the order of $10^6$ steps, as the relaxation is hindered by the initially parallel arrangement of the two arms. The long bonds pull the two arms together until they come into contact with each other. The further relaxation of the long bonds is hindered until the arms diffuse sufficiently far from a parallel arrangement.

Pictures of the three examples at different stages of the relaxation are illustrated in Fig. 2, and movies of the relaxation trajectories are available at the oxDNA youtube channel (*see* **Note 7**). In the videos of the relaxation of the six-helix bundle, the rotation of the ends as the origami adopts its overall twisted geometry is very apparent. For this origami, relief of the twist stress during the relaxation also leads to local bending that is not representative of the equilibrated state.

For a general example, visualizing the trajectory can help to show how close a system is to completing the relaxation, particularly for cases involving long bonds.

One alternative to the above scheme that has recently become available is to use the multi-resolution DNA (mrDNA) tool of Maffeo and Aksimentiev [7]. This allows modelling of DNA origami at coarser levels of detail than oxDNA. One potential advantage of this tool for relaxation is that the compute time required for relaxation may be significantly shorter when large scale motion is required due to the lower resolution of the model. The final configurations can be output to oxDNA format, and then used as starting points for a standard oxDNA molecular dynamics simulation (note, mrDNA can automatically perform an oxDNA minimization and relaxation on the configuration).

The conversion and relaxation of the three examples above should all be straightforward. However, this is not always the case. The first potential problem is in the conversion of the cadnano file. Although the underlying Python script can handle most cadnano files, it occasionally fails when less common features are present. If this occurs we recommend trying one of the other tools that can load cadnano files and output in oxDNA format, e.g. mrDNA, vHelix or Adenita.

The two most common problems in the relaxation stage are that (i) part of the DNA origami becomes irreversibly damaged due to the breaking of base pairs as a result of the internal stresses that are present in the initial structure and (ii) there are topological entanglements due to the layout of the origami structure in the design file.

To overcome the first problem one can introduce artificial "mutual traps" between those base pairs that are liable to break (*see* **Note 8**). (These traps are simply a harmonic potential in the distance between the relevant nucleotides.) This scheme would require the identification of the relevant nucleotides, and so a simpler general solution is to apply mutual traps between all pairs of nucleotides that are designed to be base-paired. A file specifying these mutual traps can be generated by one of the Python scripts associated with oxView [26] and by the Tiamat converter in tacoxDNA. The latter approach also has the potential advantage that it allows one to change the parameters of the modified backbone potential to increase the forces applied, and so enables sections connected by overstretched bonds to be brought together more quickly without having to worry about base pairs being broken.

Topological entanglements can result from the layout of the origami in the design file. This is particularly an issue for origamis designed with cadnano that involve multiple blocks or that have a structure that is not compatible with the hexagonal and square lattices available in cadnano. For example, one could represent a tube in cadnano as a flat sheet, but with crossovers between the top and bottom helices (Fig. 4). If one tries to relax such a structure in oxDNA, initially the bonds between the top and bottom helices will pass through the rest of the origami. The forces associated with these overstretched bonds will cause the sheet to buckle. If one is lucky, the sheet might coherently buckle into a C-shape, removing potential entanglements and relaxation into a tube could be successful. However, it is much more likely that some parts of the sheet will buckle more into an "S-shape", so that these long bonds still pass through the sheet (Fig. 4). In this case, relaxation will lead to a malformed structure which cannot escape from this topologically entangled state because the excluded volume in the oxDNA mode prevents nucleotides passing through each other (*see* **Note 9**).

The best way to avoid such topological problems is of course to not have them in the first place, so if one is designing an origami in cadnano that one will want to model with oxDNA, we recommend trying to organize the layout of the design so topological problems are avoided if possible, e.g. by displacing parts of the structure so that extended bonds do not pass through the rest of the origami. However, if one is working with an existing cadnano file there are a number of different approaches that have been used by different oxDNA users with new tools recently becoming available that makes this easier.

The first set of approaches involves manipulation of the structure prior to relaxation. Probably the most convenient way is to use a programme that allows a real space representation of the structure to be manipulated on screen. For example, oxView has the functionality to select origami blocks and then to translate and rotate them (*see* **Note 10**). Somewhat similarly, Adenita allows origami elements to be manipulated. In addition, oxView has a rigid-body dynamics mode that can aid the generation of sensible initial geometries. Also, manipulations of an origami structure can be performed in mrDNA via Python scripting.

An alternative way to manipulate the origami structure may be to directly edit the cadnano `.json` file prior to conversion using, for example, the online json editor available at: https://jsoneditoronline.org/, although this of course requires a certain level of understanding of the cadnano format.

The second set of approaches directs the relaxation away from topological traps. The facility in the oxDNA code to add bespoke forces to different parts of the origami is particularly useful in this regard (*See* **Note 11**). For example, for the case of the flat sheet above, if one adds forces pulling the top and bottom helices out of the plane in the same direction, this will encourage the sheet to deform into a C-shape and relax correctly (Fig. 4).

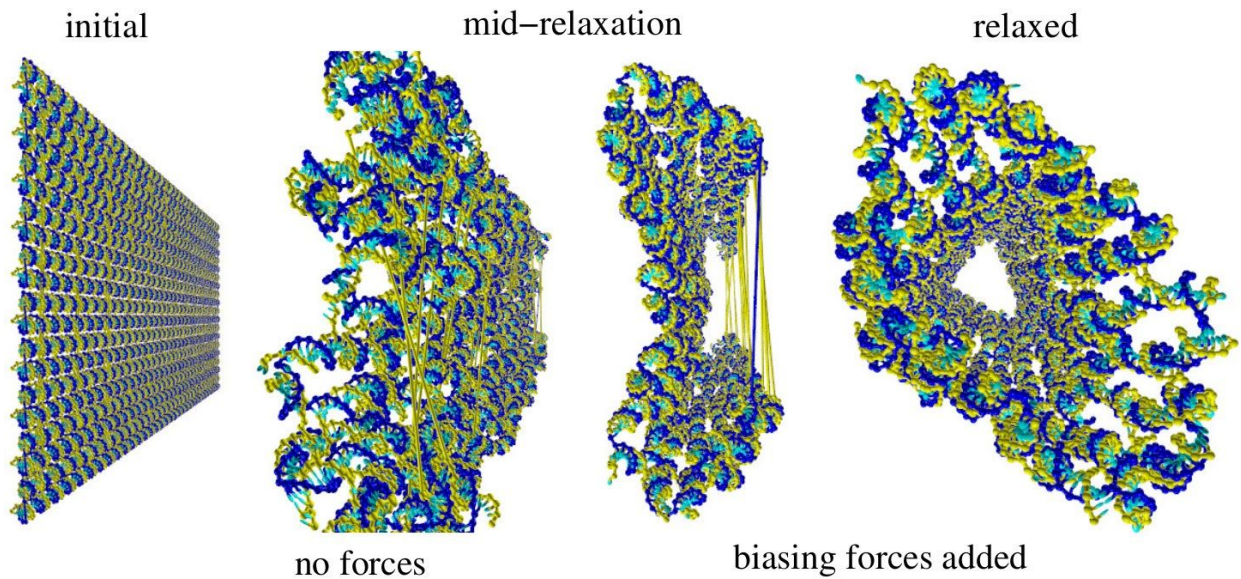

**Fig. 4** Relaxation of an origami tube that was designed in cadnano as a flat sheet but with crossovers between the top and bottom helices. If no external biasing forces are added, the long bonds will cause the sheet to buckle non-uniformly, resulting in topological entanglements on relaxation. However, if forces are added that pull the top and bottom helices to the right, the sheet deforms into a C-like configuration. Consequently, none of the long bonds pass through the rest of the structure, and relaxation to a tube configuration proceeds smoothly. Note that there are two isomers for this system depending on which surface is on the outside. The other isomer can be obtained by applying the forces in the opposite direction.

In cases with potential topological entanglements, we recommend visualization of the relaxation trajectory as spotting the entanglements in a relaxed configuration is not always straightforward.

## 3.3 Origami simulation

Once one has a sufficiently relaxed origami configuration (*see* **Note 12**), it is then just a matter of running a molecular dynamics simulation. Typically for origami we will run the simulation on a GPU for it to occur in a reasonable time frame. For each origami, example input files for this stage are provided, and the simulations are started by running the command:

    oxDNA input_sim

An important question is for how long to run the simulation. This will of course depend on what one is trying to achieve; this could range from getting a quick feel for what an origami "looks like" to comprehensively sampling the configuration space of the origami. In the latter case, the answer to this question depends on two main factors: the time required to equilibrate the system (i.e. achieve a representative starting configuration for sampling) and the time to appropriately sample the configuration space.

Even though the starting configuration needs to have been sufficiently relaxed to be a starting point for MD it may not necessarily yet be representative of the equilibrium ensemble at the temperature of interest. Therefore, an important part of any simulation is an equilibration period that allows that to be achieved. When calculating average equilibrium properties of an origami, the data from the equilibration period should of course not be included. Typically, to determine when equilibration has occurred, one looks at the behaviour of relevant properties of the system as a function of time. One obvious starting point is the energy. If we take the case of the six-helix bundle, the starting energy is actually not atypical of the ensemble. However, there are other slower degrees of freedom. For example, Fig 5(a) shows the end-to-end distance. As well as the expected twisting, the release of internal stress during the relaxation leads to considerable bending, causing the end-to-end distance in the resulting configuration to be significantly shorter than in an equilibrated configuration, and it takes at least $10^6$ steps to reach a more realistic value.

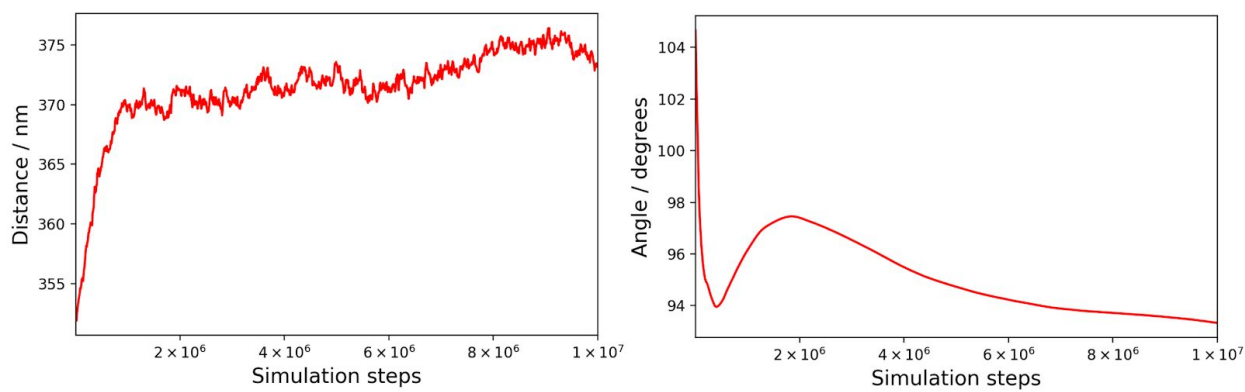

**Fig. 5** Origami properties during the simulations: (a) "end-to-end" distance (measured between points slightly in from each end where the helices do not splay out) of the six-helix bundle, (b) twist angle (measured as the angle between vectors defined by ten parallel helices in each arm) between the arms of the switch.

Somewhat similarly, an appropriate sampling time depends on the time scales associated with fluctuations in the slowest degrees of freedom (and also whether the property of interest depends on these degrees of freedom). The pointer is relatively rigid and should sample its configurational ensemble fairly rapidly. For the six-helix bundle, although again relatively rigid, if one wants to characterize the long wavelength elastic thermal fluctuations accurately one needs to use very long runs (see Ref. [13] for the unusual chiral fluctuations of this system). For the switch, the diffusive nature of the relative motion of the arms and the large length scales involved mean that fully sampling the open state will also require very long runs. As can be seen from Fig. 5, the $10^7$ step simulation runs in the examples for these two cases are insufficient to accurately calculate the probability distributions for the end-to-end distance or the inter-arm angle of the switch. In the switch, there is an even longer time scale associated with its opening and closing.

One important point is that the absolute time scales associated with coarse-grained simulations, such as those using oxDNA, should be interpreted with caution. Typically, time scales are reported using the time unit defined by the units of the basic oxDNA interactions. However, coarse-graining reduces the time scale separation between microscopic motions and diffusion times; thus, the effective time may be considerably larger, and it may be more appropriate to report times based on a mapping of diffusion times [19].

**3.4 Analysis of a simulation trajectory**

The general philosophy of the oxDNA code is to use the simulation code to produce trajectory files that can then be post-processed, rather than hard-coding lots of analysis options into the simulation code. Typically, this analysis has been done with bespoke Python scripts, such as that used to compute the twist angle of the switch in Fig. 5. To make things easier for the non-expert user, there is now a variety of general-purpose analysis tools associated with oxView that are particularly useful for analysing simulations of DNA origamis [25]. These include calculating the average structure, performing principal component analysis and clustering of configurations in a trajectory. OxView also allows the results of analyses to be overlaid on structure, for example, to identify flexible sections or points in a structure where base pairs are more likely to break due to internal stresses.

We note that the oxDNA code is also equipped with a set of "observables" that can be output during the simulation and are particularly useful when the frequency with which one needs to sample a property would otherwise necessitate excessively large trajectory files. Although mainly undocumented, details of some of the observables can be found in the `README` file accompanying the oxDNA source. Additional bespoke observables can be created through the plugin infrastructure of the code.

We also note that the UTILS directory of the oxDNA source provides a set of Python scripts that have been developed by past users. Note, however, that they are not maintained and are often undocumented, the exception to the latter being the scripts documented here: https://oxdna-utils.readthedocs.io/.

Access to examples may also be a useful source to see how particular tasks can be achieved. A number of examples come with the oxDNA source, some of which are documented at: https://dna.physics.ox.ac.uk/index.php/Category:Examples, and all the analysis scripts associated with oxView include example simulations. We also encourage users, when publishing results produced using oxDNA, to deposit relevant data, including input files and scripts used to process data. Links to these data deposits are included on the publications page of the oxDNA website: https://dna.physics.ox.ac.uk/index.php/Publications.

## 4 Notes

1. The conversion factors for the oxDNA code's simulation units are:
   1 unit of length = 0.8518 nm
   1 unit of energy = thermal energy at 3000K = $4.142 \times 10^{-20}$ J
   1 unit of temperature = 3000 K
   1 unit of force = 48.63 pN
   1 unit of mass = $5.24 \times 10^{-25}$ kg
   1 unit of time = 3.03 ps
   1 unit of force constant = 57.09 pN/nm
   1 unit of torque = 41.423 pN nM
2. Although these programmes can also output in CanDo format, the CanDo format does not explicitly specify the position of the nucleotides in single-stranded sections. If one uses the CanDo to oxDNA converter the positions this general algorithm generates for these nucleotides may not always be the most appropriate (e.g. topological entanglements may result). By contrast, the Bathe group design programmes, when outputting to `.pdb`, have specific algorithms to generate sensible positions for the nucleotides in the single-stranded sections at the vertices of the wireframe structures.
3. By default, the cadnano to oxDNA converter assigns a random sequence to the scaffold. As most origami properties have little dependence on sequence, this is expected to be unproblematic for the vast majority of applications. If there are particular reasons a specific sequence is needed (perhaps only in certain sections) the following discussion may be helpful: https://sourceforge.net/p/oxdna/discussion/general/thread/aa60af259b/
4. An alternative to the use of the minimization algorithm is to run a short Monte Carlo simulation (see, for example, the relevant input file on the tacoxDNA server).
5. To run oxDNA on a GPU the source code needs to be compiled using the flag `-DCUDA=1`, and an appropriate GPU has to be available on your machine/cluster. Note, also that there are two GPU parallelization approaches implemented in the oxDNA code. In the original, the force calculation is parallelized over particles, whereas in the second "edge-based" approach (specified by `use_edge = 1` in the input file), the parallelization is over interacting pairs of particles [37]. We recommend the edge-based approach as it is generally more computationally efficient.
6. In the relaxation MD run, it is helpful to use a tightly-coupled thermostat to more quickly remove the energy liberated by relaxation to a lower-energy configuration. In particular, we use the thermostat due to Bussi *et al.* [38] (specified by `thermostat = bussi` in the input file with the algorithm parameter `bussi_tau` controlling the tightness of the coupling). By contrast, in standard MD runs, we typically use the Andersen-like thermostat of Ref. [39] (specified by `thermostat = brownian`), with parameters designed to lead to efficient diffusion of strands.
7. The following videos are available on the oxDNA youtube channel:
   Switch relaxation (side view): https://www.youtube.com/watch?v=XpjnjYIa2N8
   Switch relaxation (top view):https://www.youtube.com/watch?v=4SPMRBIA_Rs

2xLH relaxation (side view): https://www.youtube.com/watch?v=cm2Gzrx1hVk
2xLH relaxation (end view): https://www.youtube.com/watch?v=cm2Gzrx1hVk
2xLH simulation: https://www.youtube.com/watch?v=sxW_Fz46z4Y
Tube (Fig. 4) relaxation: https://www.youtube.com/watch?v=sAjCGKe_iwA
Tube relaxation (close-up): https://www.youtube.com/watch?v=pyO5zK5ndJY

8. Mutual traps are one example of a number of types of external forces that can be applied using the native oxDNA code. To activate this feature, one must add **external_forces = true** in the input file and also supply an external force file (specified by the key **external_forces_file = <file>**). For mutual traps the format of this file is the following for each pair of nucleotides *i* and *j* interacting via a mutual trap.

```
{
type = mutual_trap
particle = <i>
ref_particle = <j>
stiff = 1.
r0 = 1.2
}
{
type = mutual_trap
particle = <j>
ref_particle = <i>
stiff = 1.
r0 = 1.2
}
```

The above provides sensible values for the bond stiffness and equilibrium internucleotide separation for the harmonic potential.

9. There is no excluded volume associated with a connection between nucleotides, so stretched bonds can pass through each other.

10. These manipulation features can also be sometimes used to aid relaxation even when there is no topological entanglement of strands. For example, if we rotate one arm of the switch by 90 degrees, when the two arms get pulled together, they will no longer collide, and relaxation occurs more rapidly (order of $10^5$ steps).

11. Adding the snippet below to an external forces file will cause a constant external force to be applied to the centre of mass of a particular nucleotide *i* in the *z*-direction:

```
{
type = string
particle = <i>
F0 = 1.0
rate = 0.0
dir = 0.0, 0.0, 1.0
}
```

The options set the magnitude of the force (in simulation units) and its direction.

12. Hopefully, the relaxation runs for the three examples are sufficiently long that the end configuration will always successfully run in the standard MD run. However, as the relaxation is a stochastic process this cannot be fully guaranteed. If the simulation does happen to fail initially, it is simply a matter of further relaxing the end configuration.

**Acknowledgements**

We are grateful for support from the EPSRC Centre for Doctoral training, Theory and Modelling in Chemical Sciences, under grant EP/L015722/1.

**Contributors**
ERIK BENSON - *Department of Physics, Clarendon Laboratory, University of Oxford, Parks Road, OX1 3PU, UK*
JOAKIM BOHLIN - *Department of Physics, Clarendon Laboratory, University of Oxford, Parks Road, OX1 3PU, UK*
JONATHAN P. K. DOYE - *Physical and Theoretical Chemistry Laboratory, Department of Chemistry, University of Oxford, South Parks Road, Oxford, OX1 3QZ, UK*
MEGAN C. ENGEL - *School of Engineering and Applied Sciences, Harvard University, 29 Oxford Street, Cambridge MA, 02138, USA*
HANNAH FOWLER - *Physical and Theoretical Chemistry Laboratory, Department of Chemistry, University of Oxford, South Parks Road, Oxford, OX1 3QZ, UK*
ARD A. LOUIS - *Rudolf Peierls Centre for Theoretical Physics, University of Oxford, Parks Road, Oxford, OX1 3PU, United Kingdom*
MICHAEL MATTHIES - *School of Molecular Sciences and Center for Molecular Design and Biomimetics, The Biodesign Institute, Arizona State University, 1001 South McAllister Avenue, Tempe, Arizona 85281, USA*
ERIK POPPLETON - *School of Molecular Sciences and Center for Molecular Design and Biomimetics, The Biodesign Institute, Arizona State University, 1001 South McAllister Avenue, Tempe, Arizona 85281, USA*
DOMEN PREŠERN - *Physical and Theoretical Chemistry Laboratory, Department of Chemistry, University of Oxford, South Parks Road, Oxford, OX1 3QZ, UK*
FLAVIO ROMANO - *Dipartimento di Scienze Molecolari e Nanosistemi, Universita Ca' Foscari, Via Torino 155, 30172 Venezia Mestre, Italy*
LORENZO ROVIGATTI - *Dipartimento di Fisica, Sapienza Universitá di Roma, Piazzale A. Moro, 2, 00185 Rome, Italy*
BENEDICT E. K. SNODIN - *Future of Humanity Institute, Department of Philosophy, University of Oxford, Littlegate House, 16-17 St Ebbe's Street, Oxford, OX1 1PT, UK*
PETR ŠULC - *School of Molecular Sciences and Center for Molecular Design and Biomimetics, The Biodesign Institute, Arizona State University, 1001 South McAllister Avenue, Tempe, Arizona 85281, USA*
JOHN S. SCHRECK - *School of Molecular Sciences and Center for Molecular Design and Biomimetics, The Biodesign Institute, Arizona State University, 1001 South McAllister Avenue, Tempe, Arizona 85281, USA*
CHAK KUI WONG - *Physical and Theoretical Chemistry Laboratory, Department of Chemistry, University of Oxford, South Parks Road, Oxford, OX1 3QZ, UK*